\documentstyle[12pt,epsf]{article}
\begin{document}
\topmargin 0pt
\oddsidemargin 1mm
\begin{titlepage}
\begin{flushright}
 OU-HET 212 \\
 hep-th/9505146 \\
 24 May 1995
\end{flushright}

\setcounter{page}{0}
\vspace{12 mm}
\begin{center}
{\Large Thermodynamics in 2+1 Dimensional \\ QED
   with Chern-Simons Term}
\end{center}
\vspace{12mm}
\begin{center}
{\large Shinya Kanemura
\footnote{e-mail: kanemu@phys.wani.osaka-u.ac.jp},
and Takao Matsushita
\footnote{e-mail: tmatsu@funpth.phys.sci.osaka-u.ac.jp}}\\
\end{center}
\vspace{3mm}
\begin{center}
{\em Department of Physics, Osaka University,\\
              Toyonaka, Osaka 560, Japan}\\
\end{center}
\vspace{5mm}

\begin{abstract}
  It is known that in the 2+1 dimensional quantum electrodynamics with
Chern-Simons term, spontaneous magnetic field induces Lorentz symmetry
breaking.
In this paper, thermodynamical characters, especially the
phase structure of this model are discussed.
To see  the behavior of the spontaneous magnetic field at finite
temperature, the effective potential in
the finite temperature system is
calculated  within the weak field
approximation and in the fermion massless limit.
We found that the spontaneous magnetic field  never vanishes at
any finite temperature.
This result doesn't change even when the chemical potential
is introduced.
 We also investigate the consistency condition and
the probability that fermion stays in
a lowest Landau level  at finite temperature.
\end{abstract}

\vspace{22mm}

\end{titlepage}

\section{Introduction}

\hspace{6mm}
In a past  decade, 2+1 dimensional gauge theory have been studied
from a lot of motivation and many interesting and remarkable
results have been  found
 such as  the chiral symmetry breaking \cite{ch1},
parity violation by radiatively inducing
Chern-Simons term \cite{pb},
and quantum Hall effects \cite{qh1,qh2}.
Two years ago, it was found that in 2+1 dimensional QED
with a bare Chern-Simons term, spontaneous magnetic
field can be stable and induce spontaneous Lorentz symmetry breaking
\cite{Ho}.

The effective potential,  calculated by Hosotani
[5] at fermion one loop level
within the weak field approximation and in the fermion massless
limit ($m_a \rightarrow 0$), is
\begin{equation}
V(B) = - \frac{e \kappa}{\pi^2}(\tan^{-1} \frac{4}{\pi} ) |B| + O(|B|^{3/2}),
\end{equation}
where $B$ is the  magnetic field.
The first term of RHS in (1)  appears not in the tree level but in
the one loop  calculation.
Of course generally, effective potential should be represented as a
Lorentz invariant form and so the order parameter should not be $B$ but
${\rm tr}F_{\mu \nu}F^{\mu \nu}$ .
Thus we can  regard (1) as the effective potential only when we consider
the model in the frame where electric field  vanishes  and only
magnetic field exists.
Looking at eq.(1), we can see immediately that true vaccum is not the point
$B=0$,
but some lowest energy state with non-zero value of $B$.
Then the spontaneous magnetic field $\langle B \rangle$ is induced
and so Lorentz symmetry is spontaneously broken.   And it can be
confirmed
that the
Nambu-Goldstone boson in this symmetry breaking is transversely
polarized gauge boson.
Further, we can read in (1) that the constant $\kappa$ which is the
coefficient of Chern-Simons term  makes  the origin  a singular
point.
Therefore it is a necessary condition  for  vanishing $\langle B \rangle$
that the first term  of RHS in (1) vanishes.

In this paper we study the thermodynamical characters of the 2+1
dimensional QED with a bare Chern-Simons term,
 especially the effective potential in the finite temperature
system \cite{ft1,ft2},
 to see the phase structure
and to investigate the
behavior of the spontaneous magnetic field in finite
temperature with chemical potential.
Our central concern is whether Lorentz  symmetry is \lq restored'
in high temperature, namely whether the spontaneous magnetic field
$\langle B \rangle$
vanishes like the case of the theory of the ferromagnetic
material
\cite{fem}.
For this aim, we calculate the coefficient of   $|B|$ in the effective
potential at finite temperature.
The calculation is carried out
in the same approximation scheme as in the ref.\cite{Ho},  namely,
within the weak field approximation
and in the fermion massless limit.
As below, we show that the spontaneous magnetic field, which exists at
$T =0$, never vanishes at any finite temperature.

In the next section, to begin with,
we start from the brief review of
this model at zero temperature for later  convenience.
And then, after introducing
temperature and chemical potential
in the ordinary manner,
we calculate the effective potential in the
finite temperature system and
discuss the behavior of the spontaneous magnetic field
at finite temperature. We also discuss the consistency  condition
and the behavior of the filling
factors at finite temperature.

\section{2+1 Dimensional QED with Chern-Simons term}

\hspace{6mm}
Consider a model described by  the Lagrangian \cite{Ho},
\begin{equation}
{\cal L} = - \frac{1}{4} F^{\mu \nu} F_{\mu \nu}
           - \frac{\kappa}{2} \epsilon^{\mu \nu \rho}
             A_{\mu}\partial_{\nu}A_{\rho}
           + \sum_{a}\bar{\psi}_a [
             \gamma_a^{\mu}(i\partial_{\mu}+q_a A_{\mu})- m_a]
             \psi_a  \,\,,
\end{equation}
\begin{equation}
F_{\mu \nu} \equiv \partial_{\mu}A_{\nu} - \partial_{\nu}A_{\nu}\,\,,
\end{equation}
where index $a$ is the label of each fermion field and the metric and
gamma matrices are defined
\begin{eqnarray}
g_{\mu \nu} &=& {\rm diag}(1,-1,-1),  \\
\gamma_a^{\mu} &\equiv& (\eta_a \sigma^3 , i \sigma^1 ,i \sigma^2 ),\\
\eta_a &\equiv& \frac{i}{2}{\rm tr}\gamma_a^0\gamma_a^1\gamma_a^2 = \pm 1 .
\end{eqnarray}
Note that there are two types of gamma matrices with $\eta_a = \pm 1$.
We call fermions
assigned $\eta_a = +1 $ type gamma matrices $\eta_+$ fermion, and
$\eta_a = -1$ type gamma matrices $\eta_-$ fermion, respectively.

The solutions  of Dirac equation for $\eta_+, q_a B > 0$
fermion are as follows,
\begin{eqnarray}
\psi (x) &=& \sum_{n=0}^{\infty} \sum_{p=-\infty}^{\infty}
             a_{np} u_{np}(x) +
             \sum_{n=1}^{\infty} \sum_{p=-\infty}^{\infty}
             b_{np}^{\dagger} w_{np}(x)  \\
E_n^a &=& \sqrt{m_a^2 + 2  |q_a B| n }
\end{eqnarray}
where $u_{np},w_{np}$ are two component spinors defined in [5], and
$a_{np},b_{np}$ are annihilation operators for fermion and
anti-fermion,
respectively. And we can get the $\eta_-$ fermion solution by
replacing
$u_{np}(x) \leftrightarrow w_{np}(x)$ in (7).
Note that there is an asymmetry of energy spectrum of
fermion  and anti-fermion at the lowest Landau level.

 Hamiltonian operator and electric
charge operator are
\begin{eqnarray}
\hat{H}_a &=& \sum_{n=0}^{\infty} \sum_p a_{np}^{\dagger}a_{np} E_n^a
           +\sum_{n=1}^{\infty} \sum_p b_{np}^{\dagger}b_{np} E_n^a
           -\sum_{n=1}^{\infty} \sum_p E_n^a , \\
\hat{Q}_a &=& \left\{
\begin{array}{c} q_a \sum_p (a_{0p}^{\dagger}a_{0p}-\frac{1}{2} )\\
                 q_a \sum_p (-b_{0p}^{\dagger}b_{0p}+\frac{1}{2})
\end{array}  \right\}   \nonumber \\
&& \;\;\;\;\;\;\; +
            q_a \sum_{n=1}^{\infty} \sum_p
            (a_{0p}^{\dagger}a_{0p} - b_{0p}^{\dagger}b_{0p})\,,\;\;
        \left\{ \begin{array}{c}  (\eta   \epsilon (B) > 0)   \\
                                  (\eta   \epsilon (B) < 0)
         \end{array} \right\}.
\end{eqnarray}

If we put the same number of $\eta_+$ and $\eta_- $
fermion ($N_f^+ = N_f^- = N_f$)
in the model to impose  chiral symmetry,
and the filling factor as $\nu^+ = 1,\,\nu^- = 0$ respectively and set
the all $q_a = e $,
the vaccum expectation value of the total charge is
\begin{eqnarray}
\langle 0\,|\,\hat{Q} \,|\,0 \rangle &=&
\sum_a \langle 0\,|\,\hat{Q}_a \,|\,0 \rangle \nonumber \\
&=& N_f^+ \langle 0\,|\,\hat{Q}_+ \,|\,0 \rangle +
N_f^- \langle 0\,|\,\hat{Q}_- \,|\,0 \rangle
= 2 N_f \langle 0\,|\,\hat{Q}_+ \,|\,0 \rangle \nonumber \\
   &=&   2 N_f     \langle 0\,|\left\{
                    e \sum_p (a_{0p}^{\dagger}a_{0p}-\frac{1}{2})
                  + e \sum_{n=1}^{\infty}
                    (a_{np}^{\dagger}a_{np} -
                     b_{np}^{\dagger}b_{np})            \right\}|\,0 \rangle
                    \nonumber \\
                &=&
                 2 N_f  e \sum_p \left( \nu^+  - \frac{1}{2} \right)
                  \, = \,
                   e \,N_p N_f \,,
\end{eqnarray}
where $\nu^{\pm}$ are the filling factor which are defined in
\begin{eqnarray}
   \nu^+ &=& \langle 0 \,| a_{0p}^{\dagger} a_{0p} |\,0 \rangle \\
   \nu^- &=& \langle 0 \,| b_{0p}^{\dagger} b_{0p} |\,0 \rangle \,,
\end{eqnarray}
and $N_p$ is the number of the degeneracy,
\begin{eqnarray}
  N_p = \frac{e B}{2 \pi} \times ({\rm area})\,.
\end{eqnarray}

Some distribution of fermion in the lowest Landau level which
satisfied  the consistency  condition
\begin{eqnarray}
\kappa = \frac{1}{2\pi}
\sum_a q_a^2 \eta_a \left( \nu_a - \frac{1}{2} \right)\,,
\end{eqnarray}
can be stable in
the presence  of spontaneous magnetic field. In our choice,
the condition (15) is
satisfied and the spontaneous magnetic field can  exist.

In this situation, the effective potential (1) is
obtained within the weak field approximation and
in the fermion massless limit.

\section{Finite Temperature System}

\hspace{6mm}
In high temperature, electron-positron pair creations  may
occur one after another and  the total particle number may be
increased,
but  the total electric charge $Q$ in this system must be conserved.
So we have to introduce the chemical potential $\mu_a$ in our
calculation as a Lagrange multiplier of conserved $Q_a$ \cite{cp}.
And the effective potential is regarded as  a thermodynamical potential $V$
which is the function of $B, T$ and $\mu_a$.

\begin{eqnarray}
  J &=& - \frac{\partial  V(B, T, \mu)}{\partial B} \\
  Q_a &=& - \frac{\partial  V(B, T, \mu)}{\partial \mu_a}
\end{eqnarray}
where $J$ is an external field coupling with $B$, and $Q_a$
is also an input parameter that we fix.
Then, we can get
\begin{eqnarray}
  B = \alpha (T, J, Q_a) B_0,
\end{eqnarray}
where $ \lim_{T \rightarrow 0} \alpha (T, J, Q_a) = 1$.
Since the external $J$ is not fixed, B remains independent variable.
However if the electric charge $Q_a$ is fixed by hand, $\mu_a$ is constrained
by eq.(17) and then  $\mu_a$ is a dependent variable and a function of
$B$ and $T$.

The $B,\,T,\,Q_a$ dependence of $\mu_a$ can
be determined from the equation
\begin{eqnarray}
Q_a = \frac{1}{\beta}
                          \frac{\partial \ln Z}{\partial \mu_a}\,.
\end{eqnarray}
where
\begin{eqnarray}
  Z &=& {\rm tr} e^{- \beta( \hat{H} - \mu \hat{Q} )} \nonumber \\
    &=& \prod_a \left( e^{\frac{1}{2}\beta \mu_a q_a} +
               e^{-\frac{1}{2}\beta \mu_a q_q}\right)^{N_p}
        \nonumber \\
    & & \times
        \prod_{n=1}^{\infty}
        \left( 1 + e^{-\beta (E_n - \mu_a q_a)} \right)^{N_p}
        \prod_{n=1}^{\infty}
        \left( 1 + e^{-\beta (E_n + \mu_a q_a)} \right)^{N_p} ,
\end{eqnarray}
where $\hat{H}=\sum \hat{H_a}$ and $\mu \hat{Q} = \sum \mu_a \hat{Q_a}$.

In our model, $q_a = e$, and $N_f^+ = N_f^- = N_f$ by chiral symmetry.
And  as setting $\nu^+ = 1, \nu^- = 0$ at zero temperature,
$Q_+ = Q_- = e N_p^0/2$,
where $N_p^0$ is $N_p$ at zero temperature.
Since the  total electric charge in this system must be conserved even
 in finite temperature, identifing
$Q ( = N_f^+ Q_+ + N_f^- Q_- )$ at finite $T$ with
$\langle 0|Q|0\rangle$
at zero temperature,
from (11), (19) and (20), we obtain
\begin{eqnarray}
  \frac{1}{2} B_0 &=& B \left\{
                      \frac{1}{2} \tanh \left( \frac{1}{2} e \beta \mu
                                                           \right)
\right.
\nonumber         \\
              & & \;\;\;\;\;\left. + \sum_{n=1}^{\infty}
                  \frac{1}{e^{\beta (E_n - e \mu )}+1 }  -
                   \sum_{n=1}^{\infty}
                  \frac{1}{e^{\beta (E_n + e \mu )}+1 }\right\}\,,
\end{eqnarray}
where $B_0$ is the magnetic field at $T=0$. Temperature dependence of
$\mu$ is shown in Figure 1. by solving this equation.
Note that in our parameter choice, $\mu \equiv  \mu^+ = \mu^-$.

We redefine the effective potential as
\begin{eqnarray}
  \Delta V(B,\,\beta,\,\mu) = V(B,\,\beta,\,\mu) - V(B=0,\,\beta,\,\mu)\,.
\end{eqnarray}
Formally introducing $T$ and $\mu$ according to Matsubara method
\cite{mm}, we
obtain
\begin{eqnarray}
  \Delta V &=& \frac{1}{2 \beta}\sum_{n=-\infty}^{\infty}
             \int \frac{d^2 \vec{p}}{(2\pi)^2} \nonumber \\
&\times&    {\rm ln} \left[
             \left(1+\Pi_0^{{\rm tot}}(p;B,\beta,\mu)\right)
               \right.  \nonumber \\
& &
   \times    \left. \left\{ \,1+\frac{1}{\vec{p}^2+p_3^2}
             \left( p_3^2\Pi_0^{{\rm tot}}(p;B,\beta,\mu)
                    +\vec{p}^2 \Pi_2^{{\rm
                        tot}}(p;B,\beta,\mu)\right)
             \right\} \right. \nonumber \\
& &         \left.
              \;\;\;\;\;\;\;
             +\frac{\left(\kappa - \Pi_1^{{\rm
                   tot}}(p;B,\beta,\mu)\right)^2}
                    {\vec{p}^2 + p_3^2}  \right] \nonumber \\
  & & - \frac{1}{2 \beta}\sum_{n=-\infty}^{\infty}
             \int \frac{d^2 \vec{p}}{(2\pi)^2}
            {\rm ln}\left\{  B \rightarrow 0\right\} \,,
\end{eqnarray}
where $p_3 \equiv 2 n \pi / \beta$
, and $\Pi_0^{{\rm tot}}, \Pi_1^{{\rm tot}}, \Pi_2^{{\rm tot}}$
are three independent  parts of
the self energies in the one particle irreducible (1PI) gauge boson
two point functions \cite{Ho} and  will be defined in the next section.

The coefficient of  $|B|$ is obtained as the total derivative of
$\Delta V$ by $B$ at
$B=0$. To calculate this, it is convenient to divide this into two parts,
\begin{eqnarray}
  \left. \frac{d \Delta V}{d B} \right|_{B=0,\mu=\mu_0(T)}
   =   \Delta V_a '
                           +  \Delta V_b '\,,
\end{eqnarray}
where
\begin{eqnarray}
 \Delta V_a '
                          &=& \left.
                              \frac{\partial \Delta V}{\partial B}
                              \right|_{
                                      B=0,
                                     \mu=\mu_0(T)    }\,,     \\
 \Delta V_b '                           &=& \left.
                              \frac{\partial \Delta V}{\partial \mu}\,\,
                              \frac{\partial \mu}{\partial B}
                              \right|_{  B=0,
                                     \mu=\mu_0(T)       } \,.
\end{eqnarray}
In (24) etc., $\mu_0(T)$ denotes the value $\mu(B=0,T)$ which
is obtained  from eq.(21).
We can see that $\mu(B=0,T) = 0$ at any temperature as below.

Since $d \Delta V/dB$ is continuous function of
$T,\mu$ for  $B \geq 0$,
\begin{eqnarray}
 \alpha (T, J=0, Q) \neq 0\;\;(T \leq T_c),
\end{eqnarray}
where $\alpha (T\,,J\,,Q)$ is defined in (18).
In eq.(27), $T_c$ is the critical temperature
where the spontaneous magnetic field vanishes and
that is just what we want to know.
Thus when $T \leq T_c$, $B = 0$ means
$B_0 = 0$.
Then, from eq.(21), we get $\mu(B=0, T) = 0$.
As far as the coefficient of  $B$  calculated  in this
way  is not zero, we can consider $T_c$ as the higher value.

{}From now, we consider (24) $\sim$ (26) with $\mu_0(T) = 0$.

Then $\Delta V_a '$ is formally written as
\begin{eqnarray}
 \Delta V_a'      &=&
 -\frac{\kappa}{ \beta}\sum_{n=-\infty}^{\infty}\int
                       \frac{d^2\vec{p}}{(2\pi)^2}
                \left.
       \frac{ \partial \Pi_1^{{\rm tot}}(p;B,\beta) }
                        {\partial B}
\right|_{B=0}     \nonumber \\
 & &  \;\;\;\;\; \times
       \left[ \left\{1+\Pi_0^{{\rm tot}}(p;B=0,\beta)
                       \right\}
                 \left\{ (p_3^2+\vec{p}^2) + \right. \right. \nonumber \\
& &   \;\;\;\;\;\;\;\;
       \left.
                      \left.
              p_3^2 \Pi_0^{{\rm tot}}(p;B=0,\beta)    +
              \vec{p}^2 \Pi_2^{{\rm  tot}}(p;B=0,\beta)
\right\} + \kappa^2 \right]^{-1}.
\end{eqnarray}
In eq.(28), we dropped the terms with
$\Pi_1^{{\rm tot}}$ and
with $\partial \Pi_{0,2}^{{\rm tot}}/ \partial B$
because of zero within the weak field approximation,
as calculated in the next section.

Thus we have  to calculate  only $\Pi_0^{{\rm tot}},
\Pi_1^{{\rm tot}}, \Pi_2^{{\rm tot}}$ at finite
temperature to calculate $\Delta V_a '$.

\section{Calculation of $\Pi_0^{{\rm tot}}
, \Pi_1^{{\rm tot}}, \Pi_2^{{\rm tot}}$ at Finite Temperature}

\hspace{6mm}
At zero temperature,
$\Pi_0,\Pi_1,\Pi_2$ are defined from the 1PI gauge boson two point
functions $\Gamma^{\mu\nu}$ \cite{Ho},
\begin{eqnarray}
  \Gamma^{00}_a &=& \vec{p}^2 \Pi_0^a \nonumber \\
  \Gamma^{0j}_a &=& p^0 p^j \Pi_0^a
                - i \epsilon^{jk} p_k \Pi_1^a \nonumber \\
  \Gamma^{i0}_a &=& p^0 p^i \Pi_0^a
                + i \epsilon^{ik} p_k \Pi_1^a  \\
  \Gamma^{ij}_a &=& \delta^{ij}(p^0)^2 \Pi_0^a
                + i \epsilon^{ij} p^0 \Pi_1^a
                - ( \vec{p}^2 \delta^{ij} - p^i p^j ) \Pi_2^a \nonumber
\end{eqnarray}
and
\begin{eqnarray}
  \Pi^{{\rm tot}}_i = \sum_a  \Pi_i^a =
              N_f \left( \Pi^{\nu =0}_i + \Pi^{\nu=1}_i \right) \;\;
  ( i = 0, 1, 2 )\,.
\end{eqnarray}

If we take the weak field approximation,
\begin{eqnarray}
  \Gamma^{\mu \nu} = \Gamma^{\mu \nu (0)}(B) + O (B^2),
\end{eqnarray}
\begin{eqnarray}
\Gamma^{\mu \nu (0)}_{\nu=0}(p)
           &\equiv& i\,q^2 \int \frac{d^3k}{(2\pi)^3}
                               {\rm tr} \left[ \gamma^{\mu}S_0^{(0)} (k)
                                               \gamma^{\nu}S_0^{(0)}
                                               (k-p)
\right] \\
\Gamma^{\mu \nu (0)}_{\nu=1}(p,\,B) &=& \Gamma^{\mu \nu (0)}_{\nu=0}(p)
                                    +  \delta \Gamma^{\mu \nu (0)}(p,\,B), \\
\delta \Gamma^{\mu \nu (0)}(p,\,B)
     &\equiv& i\,q^2 \int \frac{d^3k}{(2\pi)^3}
               \left\{ {\rm tr} \left[ \gamma^{\mu}S_0^{(0)} (k)
                                        \gamma^{\nu} f(k-p)
                                      \right]+ \right. \nonumber \\
      & & \;\;\;\;\;\;\;\;\;\;\;\;\;\;\;\;\;\;\;\;\;\;\;
                \left. {\rm tr} \left[ \gamma^{\mu}f(k)
                                       \gamma^{\nu}S_0^{(0)} (k-p)
                                      \right]           \right\}
\end{eqnarray}
where
\begin{eqnarray}
S_0^{(0)}(k) \equiv \eta \,
        \frac{ k^\mu \gamma_\mu}{ k^2 - m^2 + i \epsilon} \,,\;
f(k) \equiv 2 \pi i\, e^{- \vec{k}^2 l^2}
               \delta ( k_0 - m )( I + \sigma^3 ) \,,
\end{eqnarray}
and $l^2 \equiv 1/|qB|$.

Now in the finite temperature system, since integral for  time
component
changes to the modes summation,  we don't have any ultraviolet
singularities.
So  we can take the weak field approximation here
to calculate $\Gamma^{\mu\nu}_{\nu =0,1}$ up to $B$.
Thus we get the finite temperature versions of
$\Pi^{{\rm tot}}_0, \Pi^{{\rm tot}}_1, \Pi^{{\rm tot}}_2$.
\begin{eqnarray}
\Pi_0^{\nu=0}(p\,;\,\beta,\,B)
         &=& -\frac{1}{\beta}\,\frac{q^2}{\vec{p}^2}
              \sum_{n'} \int \frac{d^2 \vec{k}}{(2\pi)^2}\, \nonumber \\
         &\times&  \frac{ 2\, \left[ m^2 + k_0 (k_0 - p_0) + \vec{k}^2 -
                \vec{p} \cdot \vec{k} \right]}
                   {\left( k_0^2 - \vec{k}^2 - m^2 \right)
                    \left[(k_0 - p_0)^2 - (\vec{k} - \vec{p})^2
                          - m^2 \right]}+O(B^2),    \\
\Pi_0^{\nu=1}(p\,;\,\beta,\,B)
       &=&  \Pi_0^{\nu=0}(p\,;\,\beta,\,B)\, \nonumber \\
       & &     - \frac{4\pi\, q^2}{\vec{p}^2}
              \frac{1}{\beta} \frac{1}{(2\pi)^2}
              \left\{(2 m + p_0) \int d^2 \vec{k}
              \frac{e^{-(\vec{k}-\vec{p})^2 l^2}}
                   {p_0^2 + 2 m p_0 - \vec{k}^2 +
                        i \epsilon} \right.\nonumber       \\
      & &
             \;\;\;\;\;\;\;\;\;\;\;\;
               + \left.(2 m - p_0) \int d^2 \vec{k}
              \frac{e^{-(\vec{k} + \vec{p})^2 l^2}}
                   {p_0^2 - 2 m p_0 - \vec{k}^2 + i \epsilon} \right\},
                   \\
\Pi_1^{\nu=0}(p\,;\,\beta,\,B)
         &=& + \frac{1}{\beta}\,\frac{2\, q^2}{\vec{p}^2}
              \sum_{n'} \int \frac{d^2 \vec{k}}{(2\pi)^2}\, \nonumber \\
         &\times&  \frac{  m \vec{p}^2 - i (p_0 - 2 k_0)
                             (k_1p_2 -k_2 p_1) }
                   {\left( k_0^2 - \vec{k}^2 - m^2 \right)
                    \left[(k_0 - p_0)^2 - (\vec{k} - \vec{p})^2
                          - m^2 \right]}+O(B^2),   \\
\Pi_1^{\nu=1}(p\,;\,\beta,\,B)
      &=& \; \Pi_1^{\nu=0}(p\,;\,\beta,\,B)
 \nonumber     \\
      & &    + \frac{4 \pi \,i\, q^2}{\vec{p}^2}
           \frac{1}{\beta}
           \sum_{n'} \int \frac{d^2\vec{k}}{(2\pi)^2}\,
            \nonumber    \\
& &  \times \left[ \frac{(ip_1 - p_2)k_2 - (p_1 + i p_2)k_1}
          {k_0^2 - \vec{k}^2 -m^2 + i \epsilon}  \,
     \delta (k_0 - p_0 - m) e^{-(\vec{k} - \vec{p})^2 l^2}\right.
           \nonumber \\
& &  \left. + \frac{-\vec{p}^2+(p_1-ip_2)k_1+(ip_1+p_2)k_2}
            {(k_0-p_0)^2-(\vec{k} - \vec{p})^2 -m^2 + i\epsilon} \,
      \delta (k_0 - m) e^{- \vec{k}^2 l^2}\right],  \\
\Pi_2^{\nu=0}(p\,;\,\beta,\,B)
  &=& \; - \frac{4 q^2}{(\vec{p}^2)^2}
                                         \frac{1}{\beta}
                \sum_{n'} \int \frac{d^2\vec{k}}{(2\pi)^2} \nonumber \\
  & & \times \frac{p_0^2 \left[ m^2  + k_0(k_0-p_0) + \vec{k}^2
                           - \vec{p} \cdot \vec{k} \right]
                + \vec{p}^2 \left[ m^2 - k_0(k_0 - p_0) \right]}
             {\left( k_0^2 - \vec{k}^2 - m^2 \right)
              \left[ (k_0-p_0)^2 - (\vec{k} - \vec{p})^2 - m^2
                                                         \right]}
        \nonumber \\
  & &   \,\,\, + O(B^2),  \\
  \Pi_2^{\nu=1}(\,p\,;\,\beta,\,B)  &=&
  \Pi_2^{\nu=0}(\,p\,;\,\beta,\,B)\,,
\end{eqnarray}
where
\begin{eqnarray}
  k_0 \equiv i k_3 = i\,\frac{2 n' + 1}{\beta} \,\pi,\,
  p_0 \equiv i p_3  = i\,\frac{2 n}{\beta}\, \pi,\,(n,\,n'\,:\,{\rm integer})
\end{eqnarray}
Some parts of fermion modes sum can be written in the complex integral
forms by the useful of mathematical formulae \cite{ft1}.

After performing  summation and complex integral, and taking the
fermion massless limit, we finally obtain $\Pi^{{\rm tot}}$s at finite
temperature,

\begin{eqnarray}
\Pi_0^{{\rm tot}}(p\,;\,B\,,\,\beta)
&=& \frac{\pi \kappa}{4} (\vec{p}^2+p_3^2)^{-1/2} +
  A_1(p\,,\,\beta) + O(B^2) \\
\Pi_1^{{\rm tot}}(p\,;\,B\,,\,\beta)
&=& 2 \kappa e |B| \frac{1}{\vec{p}^2+p_3^2} + O(B^2) \\
 \vec{p}^2 \Pi_2^{{\rm tot}}(p\,;\,B\,,\,\beta)
&=& - p_3^2\Pi_0^{{\rm tot}}(p\,;\,B\,,\,\beta)
\nonumber  \\
& & + (\vec{p}^2+p_3^2)^{1/2} \frac{\pi \kappa}{4}
   - A_2(p\,,\,\beta) + O(B^2)\,,
\end{eqnarray}
where $A_1,A_2$ are the parts of temperature effect
from fermion loops.

\begin{eqnarray}
& & A_1(p\,,\,\beta)
  = \frac{2\kappa}{\vec{p}^2}
    \int_0^{\infty} d k
    \frac{1}{e^{\beta k}+1} \nonumber \\
& &\;\;\;\;\;\; \times
\left\{ 1 -  \left[
\frac{\sqrt{(\vec{p}^2 + p_3^2 - 4k^2)^2 + 16k^2p_3^2}
      + \vec{p}^2 + p_3^2 - 4k^2}
     {2(\vec{p}^2+p_3^2) }
\right]^{\frac{1}{2}} \right\}  \\
&&A_2(p\,,\,\beta)\
 = 2\kappa \frac{p_3^2}{\vec{p}^2}
    \int_0^{\infty} d k
    \frac{1}{e^{\beta k}+1} \nonumber \\
&& \;\;\;\;\;\;\;\;\;\;
    + \,\kappa \,\frac{(\vec{p}^2 + p_3^2)^{1/2}}
                 {\vec{p}^2}
     \int_0^{\infty} d k
    \frac{1}{e^{\beta k}+1} \nonumber \\
&& \;\;\;\;\;\;\;\;\;
      \times \frac{(*)} {\sqrt{ (\vec{p}^2 + p_3^2 - 4k^2)^2
                                     + 16 k^2 p_3^2}}
\end{eqnarray}
where
\begin{eqnarray}
(*) &\equiv& \sqrt{2}\,
\left\{ (4 k^2 - p_3^2)
        \left[\sqrt{(\vec{p}^2+p_3^2-4k^2)^2 + 16 k^2 p_3^2
     }+ \vec{p}^2+p_3^2-4k^2
                \right]^{1/2}  \right. \nonumber \\
& & \; \left. - 4 k p_3 \left[ \sqrt{ (\vec{p}^2 + p_3^2 - 4 k^2)^2
                                              +16 k^2 p_3^2 } -
                        (\vec{p}^2 + p_3^2 - 4 k^2)
                        \right]^{1/2}    \right\}.
\end{eqnarray}
In (43) $\sim$ (45), if we take the limit $T=0,\,(\beta \rightarrow
\infty)$,  it can be easily confirmed that $A_1,A_2$
vanish and only the first term remain
in  each equation  within weak field approximation
and that our calculation  reproduces the result in
ref.\cite{Ho}.

\section{Effective Potential with $T$ and $\mu$}

\hspace{6mm}
To know the behavior of the spontaneous magnetic field
in the finite temperature system,
we have to investigate the effective potential and
have to solve the gap equation generally.
However to our aim seeing whether
spontaneous magnetic field vanishes or not at finite temperature,
we have only to know the behavior
of the effective potential at neighborhood of $B=+0$,
namely the behavior of the coefficient of  $|B|$
as mentioned in previous section.

With the equations (43), (44), and (45) , we finally obtain
after straightfoward calculation from eq.(28),
\begin{eqnarray}
 \Delta V_a'
= -\frac{e \kappa}{2 \pi^2} \alpha \sum_{n=-\infty}^{\infty}
     \int_0^{\infty} d x \frac{x^5}{F_n(x\,,\,\alpha)},
\end{eqnarray}
where
\begin{eqnarray}
\alpha \equiv \frac{2 \pi T}{\kappa}\,,\;\; x \equiv \frac{p}{\kappa} \,,
\end{eqnarray}
\begin{eqnarray}
F_n(x\,,\,\alpha)
&\equiv&  \left[ x^2 \left(x^2+(\alpha n)^2 \right)
              +\frac{\pi}{4} x^2 \sqrt{x^2 + (\alpha n)^2}
             + \tilde{A}_1(x,\,\alpha,\,n)  \right]    \nonumber    \\
& & \;\times
     \left[x^2 \left( x^2 + (\alpha n)^2 \right)
            + \frac{\pi}{4} x^2 \sqrt{x^2 + (\alpha n)^2}
              - \tilde{A}_2(x,\,\alpha,\,n) \right] \nonumber \\
& &     + x^4 \left( x^2 + (\alpha n)^2 \right)\,,
\end{eqnarray}
and,
\begin{eqnarray}
& &\tilde{A}_1(x\,,\,\alpha\,,\,n) \nonumber \\
& &\equiv \alpha \frac{\log 2}{\pi}
         \left( x^2 + (\alpha n)^2 \right)
- \frac{\alpha}{\pi}\left( \frac{x^2+(\alpha n)^2}{2}
                                             \right)^{\frac{1}{2}}
        \int_0^{\infty} dy \frac{1}{ e^y + 1 }  \nonumber \\
& & \times
        \left\{ \sqrt{\left(x^2+(\alpha n)^2-
                              \frac{\alpha^2}{\pi^2} y^2 \right)^2
                         +\frac{4}{ \pi^2 }  \alpha^4 n^2  y^2 }
                  + x^2 + (\alpha n)^2 -\frac{\alpha^2}{\pi^2} y^2
                                                 \right\}^{\frac{1}{2}},
\end{eqnarray}
\begin{eqnarray}
& &\tilde{A}_2(x\,,\,\alpha\,,\,n) \nonumber \\
& &\equiv \alpha^3 n^2 \frac{\log 2}{\pi}
         + \alpha^3 \left( \frac{x^2 + (\alpha n)^2}{2}\right)^{\frac{1}{2}}
           \frac{1}{\pi}\int_0^{\infty} d y
           \frac{1}{e^y +1}  \nonumber \\
& &\times \left\{ \left( \frac{y^2}{\pi^2} - n^2 \right)\,
                 \left[ \frac{ \sqrt{ \left( x^2 + (\alpha n)^2 -
                                             \frac{1}{\pi^2} \alpha^2
                                             y^2
                                                            \right)^2
                               + \frac{4}{\pi^2} \alpha^4 n^2 y^2}
                              +x^2 +(\alpha n)^2 -
                                \frac{1}{\pi^2} y^2 \alpha^2 }
                             {\left(x^2+(\alpha n)^2 -\frac{1}{\pi^2}
                                    \alpha^2  y^2 \right)^2
                              +\frac{4}{\pi^2}\alpha^4 y^2 n^2}
                                            \right]^{\frac{1}{2}}\right.
\nonumber \\
& & \left. -\frac{2}{\pi}n y
\left[ \frac{\sqrt{ \left( x^2 + (\alpha n)^2 -
                                             \frac{1}{\pi^2} \alpha^2
                                             y^2
                                                            \right)^2
                               + \frac{4}{\pi^2} \alpha^4 n^2 y^2}
       -\left( x^2 +(\alpha n)^2 -
                                \frac{1}{\pi^2} y^2 \alpha^2 \right)     }
 {\left(x^2+(\alpha n)^2 -\frac{1}{\pi^2}
                                    \alpha^2  y^2 \right)^2
                              +\frac{4}{\pi^2}\alpha^4 y^2 n^2}
                                       \right]^{\frac{1}{2}}\right\}\,.
\end{eqnarray}

In eq.(49), since we can analytically show in high temperature limit
and in low temperature limit and numerically for all temperature
 that
\begin{eqnarray}
  F_n(x,\,\alpha) > 0,\;\forall \,x,\,\alpha,\,n,
\end{eqnarray}
we confirm that $\Delta V_a '$ is always negative.
\begin{eqnarray}
  \Delta V_a'  < 0,\;\forall \,\,\alpha .
\end{eqnarray}
In high temperature region, to know the asymptotic behavior of
$\Delta V_a'$, it is convenient to divide $\Delta V_a'$ into
the zero mode part and the another part.
Permuting  integral variable $x$ to $\alpha u$ in the zero mode
part and to $\alpha n \sinh v$ in the another part,
we can easily see that the each part have
$\alpha^{-1}$ dependence asymptotically.
Thus,
\begin{eqnarray}
  \Delta V_a' \propto - \, \frac{1}{T},\;\;(T \rightarrow \infty).
\end{eqnarray}
And we have analytically confirmed that in $T \rightarrow 0$, eq.(49)
reproduces the result in ref.\cite{Ho} using Euler - Maclaurin's
mathematical formula \cite{NB},
\begin{eqnarray}
  \lim_{T \rightarrow 0} \Delta V_a' = - \frac{e \kappa}{\pi^2}
          \tan^{-1} \frac{4}{\pi}.
\end{eqnarray}

We also numerically calculated $\Delta V_a'$, (see Figure 2).
We can see from Figure 2 that, except for near $T=0$,
$\Delta V_a'(T)$ is monotonously increasing to zero
having the behavior $\sim -1/T$.
This results seem to say that the spontaneous magnetic field would
not vanish even in high temperature region,
say $T_c = \infty$, if the second term,
$\Delta V_b'$ wouldn't have any effect in $d \Delta V / d B$ .

Of course, nextly $\Delta V_b'$ have also to be considered in
our study.
However,  we can know the qualitative but essential behavior of
$\Delta V_b '$ without direct calculation as below.

At first, from eq.(21), we can  analytically check that in neighborhood
of $B=0$, $\partial \mu / \partial B$ is positive for all temperature
and its $B$ dependence is at most order minus one,
\begin{eqnarray}
  \frac{\partial \mu}{\partial B} \propto
   \frac{1}{B} + O (1) > 0, \;\;( B \sim + 0,\;\forall \,\, T ).
\end{eqnarray}
And we can  see
\begin{eqnarray}
  \frac{\partial \mu}{\partial B} \sim
 \left\{ \begin{array}{cl}
            \frac{1}{\sqrt{B}} & (T \rightarrow 0) \\
            \frac{1}{T}        & (T \rightarrow \infty)
         \end{array} \right.
\end{eqnarray}
from eq.(21).
Further,  since from eq.(17)
\begin{eqnarray}
  \frac{\partial \Delta V}{\partial \mu} &=& - \frac{  Q (B,\,T\,\mu)
      - Q (B=0,\,T,\,\mu) }{2 N_f}  \nonumber
\\            &=&
  - \frac{e^2}{4\pi} \frac{B}{\alpha (T)} < 0 ,\
\end{eqnarray}
After all, $\Delta V_b'$ is negative for all $T$ and
\begin{eqnarray}
  \Delta V_b' \rightarrow  \, \, - 0 ,\;\;(T \rightarrow 0,\, \infty)
\end{eqnarray}

Therefore, $\Delta V_b'$ have  essentially no any effects
in the phase structure of the order parameter $B$, especially
in the  both limit $T \rightarrow 0 ,\; \infty $,
$\Delta  V_b '$ is to be zero (Figure 2).

\section{Consistency  Condition at Finite Temperature}

\hspace{6mm}
In the case $T=0$, as we are setting up the filling factor
$\nu^+ = 1,\,\nu^- = 0\,( {\rm and}\; N_f^+ = N_f^- = N_f$
by chiral symmetry ),
the lowest Landau level  degenerating to $x^1$
direction is filled by $\eta_+$ fermions and they keep the generating
spontaneous magnetic field $B$ and there are no $\eta_-$ fermions.
Then the consistency condition (15) is of course satisfied,
\begin{eqnarray}
  \kappa = \frac{e^2}{2\pi} N_f.
\end{eqnarray}
This condition is the sufficient condition for having
the spontaneous
magnetic field $B_0$ stably as mentioned before.

In finite temperature system,  the consistency  condition (15)
have to be rewitten as follows,
\begin{eqnarray}
  \kappa = \frac{e^2}{2\pi}
           \sum_a \eta_a
           \left( \langle \nu_a \rangle - \frac{1}{2} \right) + 2 N_f e
           \langle \sum_{n=1}^{\infty} \sum_p
            \left( a_{np}^{\dagger} a_{np} -
                              b_{np}^{\dagger} b_{np}  \right)
          \rangle / B,
\end{eqnarray}
where
\begin{eqnarray}
  \langle x \rangle  \equiv  \frac{{\rm tr} x e^{- \beta (\hat{H}- \mu
      \hat{Q})} }{{\rm tr} e^{- \beta (\hat{H}- \mu
      \hat{Q})}}.
\end{eqnarray}

In our model, we have just seen that spontaneous
magnetic field never vanishes and $T_c$ has gone  to infinity.
In finite temperature, $\eta_+$ fermions which are in the ground states (
the lowest Landau level ) may be
gradually excited to higher states.
Then, naively $\langle \nu^+ \rangle$ seems to decrease as temperature grows.
In fact, as the filling factor operators of $\eta_{\pm}$ fermion are
defined at finite temperature,
\begin{eqnarray}
  \hat{\nu}^+ = \frac{1}{N_p} \sum_p a_{0p}^{\dagger} a_{0p}\,,\;
  \hat{\nu}^- = \frac{1}{N_p} \sum_p b_{0p}^{\dagger} b_{0p}\,,
\end{eqnarray}
we can easily calculate
 ensemble averages as follows,
\begin{eqnarray}
\langle \nu^+ \rangle = \frac{1}{e^{- e \beta \mu} + 1}\,,\;
\langle \nu^- \rangle = \frac{1}{e^{+ e \beta \mu} + 1} \,.
\end{eqnarray}
Clearly, $\langle \nu^+ \rangle \,(\,, \langle \nu^- \rangle)$ is
monotonously decreasing (, increasing) functions
of  $T$ (see Figure 3.)
and in high temperature limit,
$\langle \nu^{\pm} \rangle \rightarrow 1/2$.
This means,
the probability that $\eta_+$ fermion exists in a ground state
is multiplied by no any statistical weight in high temperature limit.
The second term in eq.(63) comes  from the charge which is shared by
the excited fermion and anti-fermion  and this can be got from eq.(21),
\begin{eqnarray}
        \frac{N_f e^2}{\pi} \left\{  \sum_{n=1}^{\infty}
                  \frac{1}{e^{\beta (E_n - e \mu )}+1 }  -
                   \sum_{n=1}^{\infty}
                  \frac{1}{e^{\beta (E_n + e \mu )}+1 } \right\}.
\end{eqnarray}

\section{Conclusion}

\hspace{6mm}
In this paper, we have examined the thermodynamical characters in 2+1
dimensional QED with Chern-Simons term, which the vaccum can be
spontaneously magnetized by some preferable fermion distribution
which satisfies the consistency condition.
The behavior of the effective potential at finite temperature is
calculated at  one loop level
within the weak field approximation and
 in  the fermion massless limit.
The coefficient of  $|B|$ in the effective potential existing at zero
temperature does not vanish even  at  finite temperature.
Thus  $B = 0$
point remains being singular and the lower vaccum at some $B \neq 0$
point  exists.

Therefore we conclude that in our model the spontaneous
magnetic field  never vanishes at any finite temperature
 differently from the case of the theory  of
the four dimensional ferromagnetic material.
We also discussed the consistency condition for having the
spontaneous magnetic field at finite temperature and
calculated the ensemble average of the filling factor - the
probability that fermion stays in a ground state as the
function of temperature and chemical potential.

Our results that  the critical temperature  $T_c$ is infinite
may seemingly look like  strange
if readers consider them
in a realistic system, such as the 2+1 dimensional system in the
3+1 dimensional world.
In this case,  the hopping to 3+1 dimension must occur at some  finite
critical temperature  and
our considering situation must break down.
However, as far as we stand on the 2+1 dimensional world, our results are
consistent.

\vspace{1cm}
\noindent
{\large \em Acknowledgements}

Authors would like to thank H. Suzuki for useful suggestion and
conversation, and  Prof. K. Higashijima for valuable discussions
and for reading this manuscript.

\newpage

\pagebreak[4]
\topmargin 0pt
\oddsidemargin 5mm
\centerline{\bf FIGURE CAPTIONS}

\begin{description}
\item[{\bf Fig.1}:]
 The behavior of the chemical potential $\mu$ as a function of
  $B,T$.  The solid line is the curve when $e B/\kappa^2 = 1/2$.
 The dashed line is the curve when $e B/\kappa^2  = 1/4$.
\item[{\bf Fig.2}:]
 $T$ dependence of $\Delta V_a'$ and $\Delta V_a' + \Delta V_b'$.
 The solid line which represents $\Delta V_a'$ is monotonously increasing
 up to zero except for near $T=0$.
 At the point $A$,  $\Delta V_a'$ has the value
 of  $ - e \kappa/\pi^2  \tan^{-1} 4/\pi$ as in the eq.(57).
 The dashed line which approximately represents
 $\Delta V_a' + \Delta V_b'$
 is always below the solid line and in the limit $T \rightarrow 0, \infty$,
  $\Delta V_b'$ turn to be zero rapidly.
\item[{\bf Fig.3}:]
 The behavior of the filling factors for $\eta_+,\,\eta_-$ fermion.
 The solid line represents $\langle \nu^+ \rangle$ and the dashed
 line represents $ \langle \nu^- \rangle$.
\end{description}

\end{document}